\documentclass[runningheads]{llncs}
\usepackage{amsmath}
\usepackage{graphicx}
\usepackage{subfigure}
\usepackage{array}
\usepackage{multirow}
\usepackage{todonotes}
\usepackage{amssymb}
\newcolumntype{L}[1]{>{\raggedright\let\newline\\\arraybackslash\hspace{0pt}}m{#1}}
\newcolumntype{C}[1]{>{\centering\let\newline\\\arraybackslash\hspace{0pt}}m{#1}}
\newcolumntype{R}[1]{>{\raggedleft\let\newline\\\arraybackslash\hspace{0pt}}m{#1}}
\DeclareMathOperator*{\argmin}{arg\,min}

\begin{document}
%
\title{An amplified-target loss approach for photoreceptor~layer~segmentation in pathological~OCT~scans}
\titlerunning{Amplified-target losses for photoreceptor~layer~segmentation in OCT}
%

\author{Jos\'e Ignacio Orlando\inst{1}\thanks{equal contribution} \and
Anna Breger\inst{2}$^\star$ \and 
Hrvoje Bogunovi\'c\inst{1} \and
Sophie Riedl\inst{1} \and
Bianca S. Gerendas\inst{1} \and
Martin Ehler\inst{2} \and
Ursula Schmidt-Erfurth\inst{1}}
%

\authorrunning{Orlando \& Breger \textit{et al.}}
%

\institute{Christian Doppler Laboratory for Ophthalmic Image Analysis (OPTIMA), Department of Ophthalmology and Optometry, Medical University of Vienna, Austria \and
Department of Mathematics, University of Vienna, Austria}
%
\maketitle              

\begin{abstract}
Segmenting anatomical structures such as the photoreceptor layer in retinal optical coherence tomography (OCT) scans is challenging in pathological scenarios. Supervised deep learning models trained with standard loss functions are usually able to characterize only the most common disease appearance from a training set, resulting in suboptimal performance and poor generalization when dealing with unseen lesions. In this paper we propose to overcome this limitation by means of an augmented target loss function framework. We introduce a novel amplified-target loss that explicitly penalizes errors within the central area of the input images, based on the observation that most of the challenging disease appearance is usually located in this area. We experimentally validated our approach using a data set with OCT scans of patients with macular diseases. We observe increased performance compared to the models that use only the standard losses. Our proposed loss function strongly supports the segmentation model to better distinguish photoreceptors in highly pathological scenarios.
\end{abstract}

\section{Introduction}
\label{sec:introduction}

Supervised deep learning techniques have revolutionized the field of medical image segmentation~\cite{litjens2017survey}, particularly with fully convolutional neural network architectures such as the U-Net~\cite{ronneberger2015u}. 
To learn these networks, a loss function $L$ is optimized using gradient based approaches and backpropagation. This function is usually defined in terms of metrics that quantify the discrepancies between a trustworthy/ground truth labelling and the predicted segmentation.

In this typical framework a loss function is not explicitly tailored to aim for a specific feature in the target space. Hence, the network firstly learns the dominating characteristics of the target images in the training set, and its remaining capacity is gradually devoted to characterize other less prevalent target features. This becomes an issue when dealing with highly pathological data, where lesions or disease appearance might significantly differ between patients. To overcome this limitation, some authors proposed to train segmentation models using a linear combination of different losses such as cross-entropy and Dice~\cite{khened2019fully}. However, these metrics are still computed from the same target representation, so they do not enhance a specific target feature. In this paper we propose to extend this idea by using the framework of \textit{augmented target loss functions}, introduced in~\cite{ortho}. Rather than relying on a single or a linear combination of loss functions defined on the original prediction and target space, Breger \textit{et al.}~\cite{ortho} proposed to compute the loss on alternative representations of the predictions and targets, obtained by applying differentiable transformations $T$ that enhance specific characteristics. 

\begin{figure}
    \centering
    \includegraphics[width=0.99\textwidth]{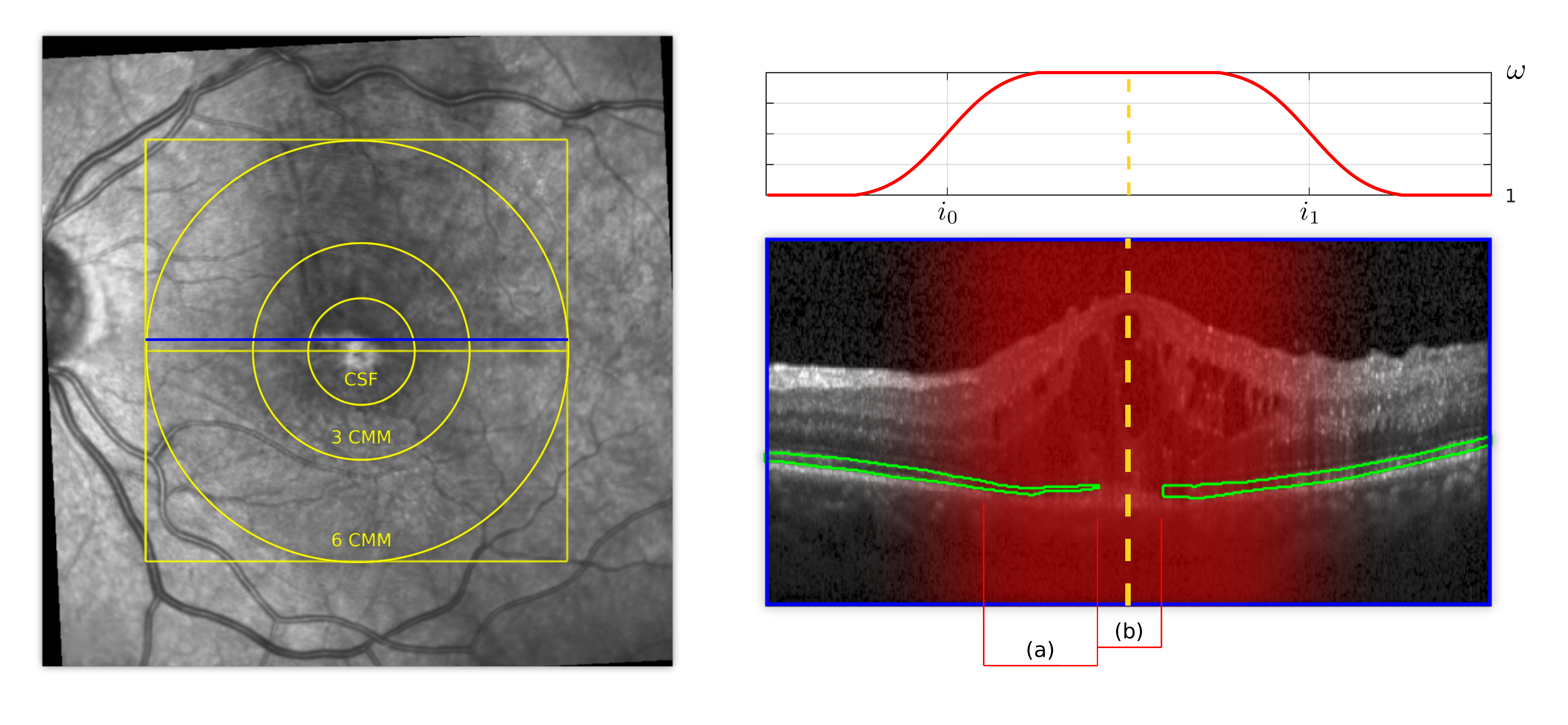}
    \caption{Left: scanning laser ophthalmoscopy (SLO) of a patient with RVO. The square indicates the area captured by the OCT volume and the rings represent the central subfield (CSF) and the 3 and 6 central millimeters (3 CMM and 6 CMM). The blue line highlights the B-scan showed in the right side. Right: CSF B-scan with photoreceptor layer annotation (green) with (a) disruptions and (b) abnormal thinning. The red heat map represents the weighting strategy applied in our loss function. The central coordinate of the image is indicated with the yellow dotted line, and a profile of the weighting strategy is illustrated on top of the B-scan. }
    \label{fig:custom-weighting}
\end{figure}

This paper focuses on the application of an augmented target loss function for photoreceptor layer segmentation in retinal optical coherence tomography (OCT) scans of patients with macular diseases. OCT is the state-of-the-art technique for imaging the retina, as it brings volumetric information through a stack of 2D scans (B-scans) at a micrometric resolution~\cite{SE2018ai}. Ophthalmic disorders such as diabetic macular edema (DME), retinal vein occlusion (RVO) and age-related macular degeneration (AMD) gradually affect photoreceptors while progressing. The abnormal accumulation of fluid due to these diseases significantly alters the retina, eventually leading to photoreceptor cell death. This last characteristic can be noticed through OCT imaging: first as a pathological thinning of the photoreceptor layer, and more lately as complete disruptions on it (Fig.~\ref{fig:custom-weighting}, right). It has been observed that these abnormalities are highly correlated with focal vision impairment~\cite{takahashi2016photoreceptor} and visual acuity loss when located at the central area of the retina~\cite{gerendas2017oct}. Hence, the automated characterization of the morphology of the photoreceptor layer is relevant for efficient quantification of functional loss. 

In this paper we build on top of the architectural innovations proposed in~\cite{orlando2019u2net} by training such a model using an augmented target loss function. Fitting the framework we introduce a novel amplified-target loss that induces further penalization to errors within the central area of the B-scans. As the most challenging pathologies are usually observed at the central area of fovea-centered OCT scans, our hypothesis is that incorporating this loss function as a kind of regularizer enforces the network to better characterize disease appearance. We validate our approach using a series of OCT scans of patients with AMD, DME and RVO. Our results empirically show that the proposed loss functions improve the performance within the central millimeters of the retina compared to using traditional losses without compromising the performance in the entire volume.

\section{Methods}
\label{sec:methods}
\subsection{Augmented target loss functions for image segmentation}
\label{subsec:variational-losses}

In a supervised learning problem we aim to learn a function $f$ with $f_\theta(x) \approx y$, where $\theta$ denotes the free parameters and $S = \{(x,y)^{(i)} \}, 1 < i < N$ is a given training set with pairs of inputs $x$ and ground truth labels $y$. In the context of image segmentation, $x$ corresponds to an input image, $y$ and $\hat{y}$ are manual and predicted segmentations and $f_\theta$ is some segmentation model (e.g. a fully convolutional neural network such as the U-Net~\cite{ronneberger2015u}). 

To adjust the weights $\theta$ from the chosen network structure $f_\theta$, a loss function $L$ is minimized using gradient based optimization. $L$ is a piecewise differentiable loss function, e.g. cross-entropy (CE) or mean squared error (MSE), that measures the pixel-wise differences between $\hat{y}$ and $y$. In standard settings no specific areas of the images are penalized more than others. Thus, the parameters $\theta$ are mostly adjusted to characterize those features from the training set that have the most impact on the overall loss. Although this might be helpful to segment healthy anatomy, in pathological scenarios the network will overfit the prevalent features unless explicit regularization is imposed during training.

Here, we propose to use the framework of augmented target (AT) loss functions, introduced in \cite{ortho}. These losses take into account prior knowledge of target characteristics via error estimation in transformed target spaces. The framework can be applied to any supervised learning problem based on loss optimization where additional information about the target data is available, provided it can be formulated as a transformation function $T$. The transformation may correspond to any piecewise differentiable function on the target space that yields a more beneficial representation of some known target characteristic.  

Following \cite{ortho}, the AT loss functions $L_\text{AT}$ is a linear
combination of losses applied to transformed targets. Its general form is: 
\begin{equation}\label{atloss}
L_\text{AT} = \sum_{j=1}^d \lambda_j \cdot L^j \big(\{T_j(y_i)\}_i^{},\{T_j(\hat{y}_i)\}_i^{} \big), 
\end{equation}
where $\lambda_j > 0$ corresponds to some weight, $T_j$ to a specific transformation and $L^j$ to some loss function, for all $j \in \{1,\dots,d\}$. 

Setting typically $T_1$ to the identity and $L^1$ to a standard loss, the additional terms in the $L_\text{AT}$ loss act as 
 amplified target information, yielding a new optimization problem:
\begin{equation}\label{eq:optimization}
    \hat{\theta} = \argmin_\theta \{ \lambda_1 \cdot L_1\big(\{y_i\}_i^{},\{\hat{y}_i\}_i^{}\big) + \sum_{j=2}^d \lambda_j \cdot L^j \big(\{T_j(y_i)\}_i^{},\{T_j(\hat{y}_i)\}_i^{}\big), 
\end{equation}
where the weights $\lambda_1$ and $\{ \lambda_j \}_{j=2}^d$ control the balance between the main loss and the regularization terms respectively.

\subsection{Amplified-target loss functions for photoreceptor layer segmentation}
\label{subsec:our-loss}

We experimentally study the AT loss function framework in the context of photoreceptor layer segmentation in pathological OCT scans. We tailor a so called \textit{amplified-target loss} in which a transformation $T$ is designed to bring an increased penalty to errors within the central area of the images. This loss is intended to incorporate the prior knowledge that abnormalities such as pathological thinnings and disruptions of the photoreceptor layer are more common in the central millimeters of the foveal area. To do so, we define a transformation $T(y_i) = W \odot y_i $, where $\odot$ denotes the Hadamard (entrywise) product, $y_i$ corresponds to the given binary targets and $W$ represents a weighting matrix that encodes a penalization weight for errors. This operation can analogously be applied to the predictions $\hat{y}_i$. Fig.~\ref{fig:custom-weighting} graphically illustrates the design of the weighting matrix $W$. Formally, we define $W = G_\sigma * V$, where $G_\sigma$ stands for a Gaussian filter with standard deviation $\sigma$. We define $V$ as:
\begin{equation} \label{eq:weighting}
    V_{j,i} := \left\{
        \begin{array}{ll}
            \omega & \qquad \text{for } i_0 < i < i_1 \text{ and all } j, \\
            1 & \qquad \text{otherwise,}
        \end{array}
        \right.
\end{equation}
 where $\omega$ denotes the maximum weight assigned to the central area and $[i_0, i_1]$ is the horizontal interval of the image that is amplified. The Gaussian filter $G_\sigma$ is used to smooth the penalization factor within the edges of the interval. 

Following the formulation in (\ref{eq:optimization}), we can then redefine our empirical risk minimization problem as
\begin{equation}
\hat{\theta} = \argmin_\theta \{ \lambda_1 \cdot L^1  \big(\{y_i\}_i^{},\{\hat{y}_i\}_i^{}\big) + \lambda_2 \cdot L^2 \big(\{ W \odot y_i \}_i^{}, \{W \odot  \hat{y}_i \}_i^{}  \big) \}, 
\end{equation}
where we choose $\lambda_1, \lambda_2  \in \mathbb{R}$ and $L^1 = L^2$ as CE or MSE losses.

\section{Experimental setup}
\label{sec:experimental-setup}

\subsection{Materials}
\label{subsec:materials}

Our method was trained and tested on an in-house data set with 53 Spectralis OCT volumes of patients suffering from DME (16), RVO (27) and AMD (10). Each image comprises $496 \times 512$ pixels per B-scan, 49 B-scans per volume. All the B-scans were manually annotated by certified readers under the supervision of a retina expert, who modified the labels when necessary to obtain ground truth segmentations. The set was randomly divided into a training, a validation and a test set, each of them with 34, 4 and 15 scans, respectively, with approximately the same distribution of diseases and percentages of disrupted columns per B-scan (or A-scans).

\subsection{Network architecture and training setup}
\label{subsec:training-setup}

We used the photoreceptor segmentation network described in~\cite{orlando2019u2net} in our experiments (note that any other architecture could be applied within our framework). We used as baselines CE and MSE comparing it to the adapted AT loss.

Every configuration was trained at a B-scan level with a batch size of 2 images, using Adam optimization and early stopping. Hence, training was stopped if the validation loss did not improve for the last 45 epochs. The learning rate was set to $\eta = 0.0001$, and divided by 2 if the validation loss was not improved during the last 15 epochs. Data augmentation was used in the form of random horizontal flippings. Binary segmentations were retrieved as in~\cite{orlando2019u2net} by thresholding the softmax scores of the photoreceptors class using the Otsu algorithm. 

\section{Results and Discussion}
\label{sec:results}

We evaluated the performance for segmenting the photoreceptor layer using the volume-wise Dice index, at the CSF, the 3 CMM, the 3-1 ring and the full volume (Fig.~\ref{fig:custom-weighting}, left). All the experiments with our AT loss functions were conducted using fixed values for $\sigma=\frac{1}{16}X$, $i_0=\frac{1}{4}X$ and $i_1=\frac{3}{4}X$ (with $X=512$ being the horizontal size of the B-scans, in pixels), without optimizing them on the validation set. Different configurations for $\omega = 2^k, k \in \{1, ..., 5\}$ and $\lambda_1$ and $\lambda_2 \in \{ 0.001, 0.01, 0.1, 1, 2, 4, 8$ \} were analyzed, and the best configuration according to Dice index on the validation set was then fixed to allow a fair comparison on the test set. From this model selection step, we observed that $\omega=8$, $\lambda_1=1$ and $\lambda_2=8$ reported the best performance for the AT loss with categorical cross-entropy (CE), and $\omega=32$, $\lambda_1=\lambda_2=1$ for the AT loss with mean square error (MSE).

\begin{figure}[t]
    \centering
    \subfigure[Cross entropy]{\includegraphics[width=0.45\textwidth]{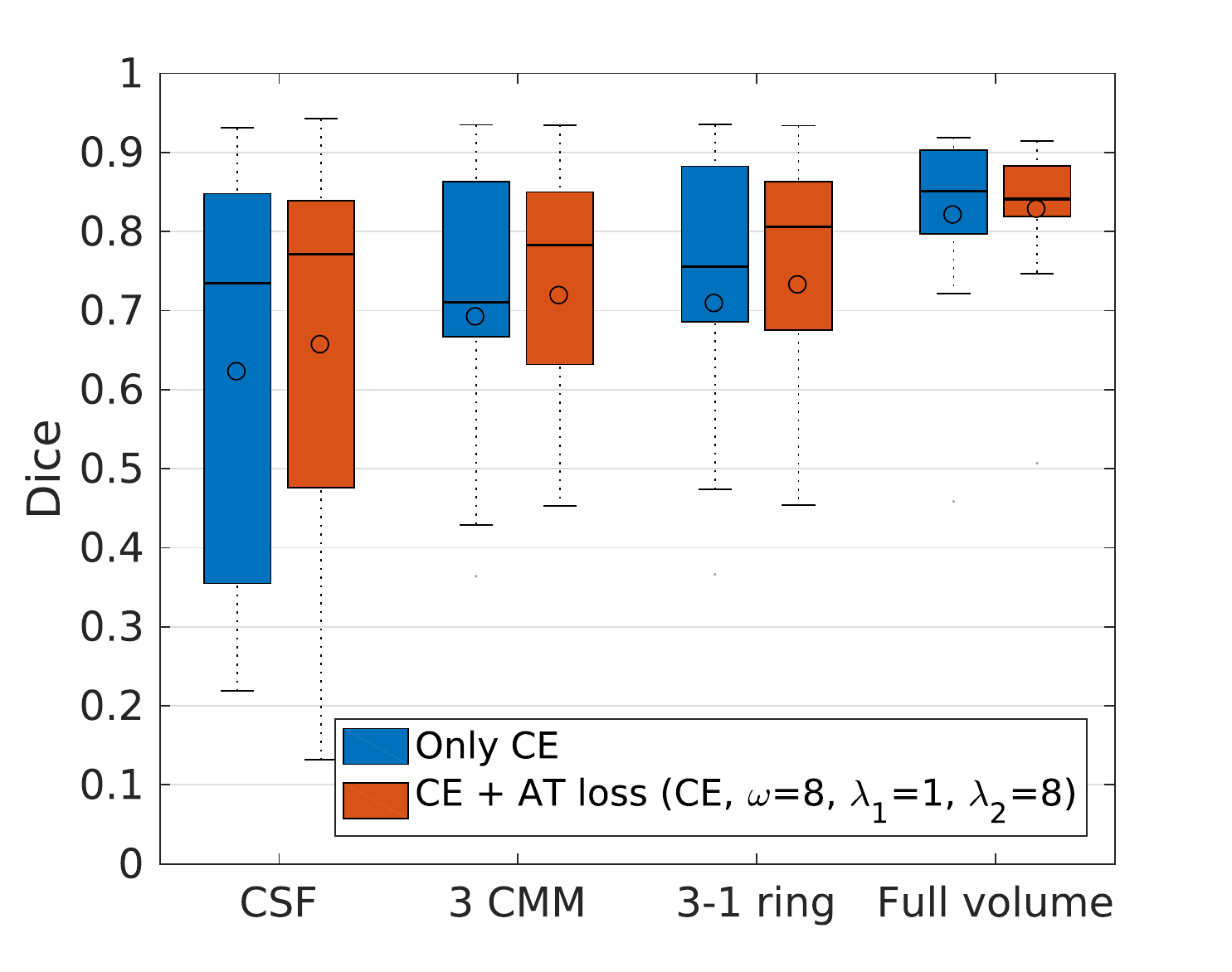}}
    \subfigure[MSE]{\includegraphics[width=0.45\textwidth]{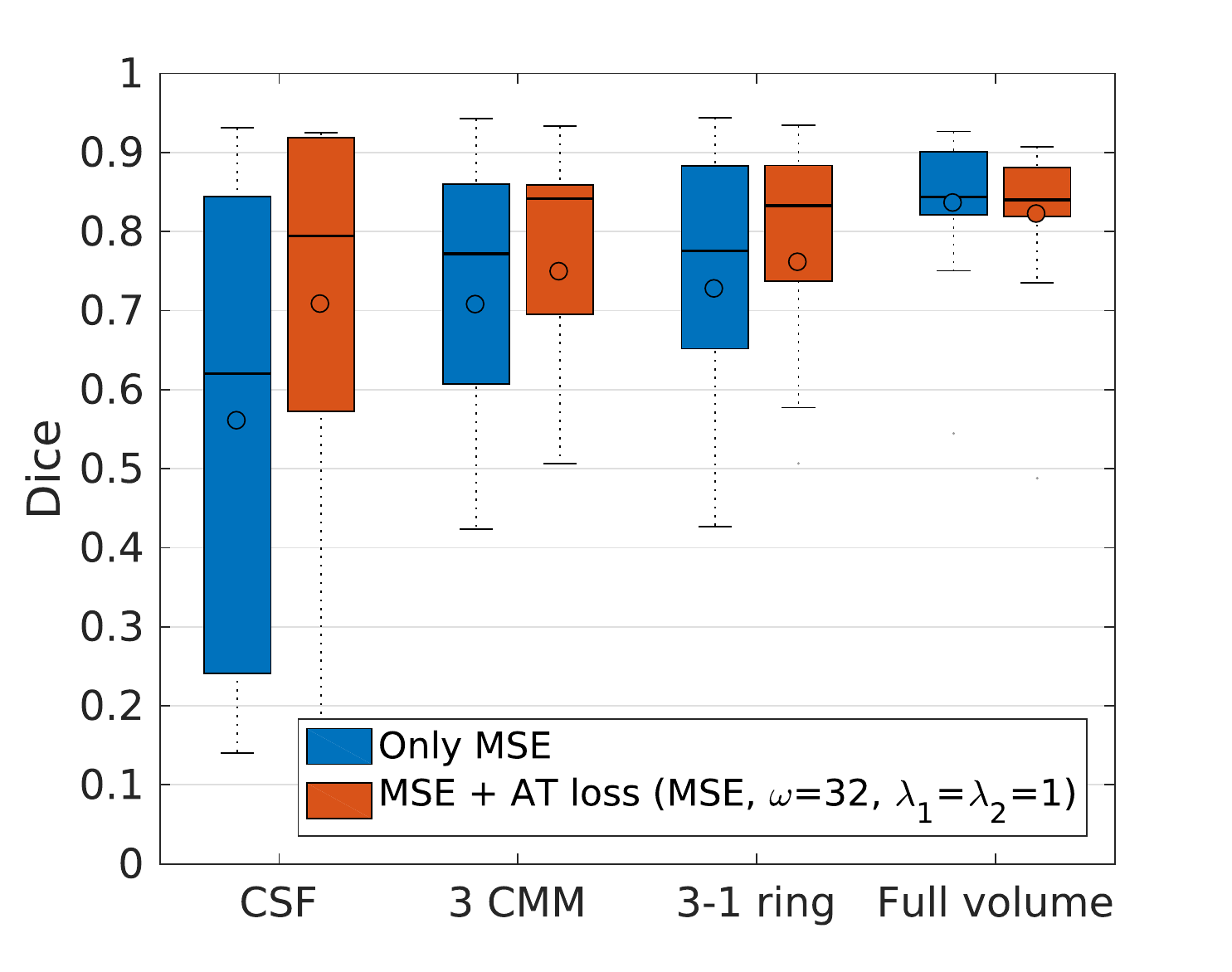}}
    \caption{Volume-wise Dice values for all the evaluated models and our proposed approach in each evaluation area. Circles indicate mean values. CSF: central subfield (1 central millimeter). 3 CMM: three central millimeter. 3-1 ring: area between CSF and 3 CMM.}
    \label{fig:quantitative-results}
\end{figure}

Fig.~\ref{fig:quantitative-results} depicts boxplots with the quantitative performance of each model on the test set, compared with their corresponding baselines trained only with CE and MSE, for each evaluation area. The mean and standard deviation values of the Dice index are presented in Table~\ref{tab:table-dice}. The incorporation of the AT loss allows to perform consistently better in all the cases, with the best results reported by the MSE loss. Statistical analysis using one-tail Wilcoxon sign-rank tests at a significance level $\alpha = 0.05$ showed that the model trained with MSE + AT loss reported significantly higher Dice values in the CSF area compared to using CE + AT loss or only MSE ($p < 0.0171$). These differences were not statistical significant with respect to the model trained with CE ($p = 0.1902$). When comparing the Dice values at the 3-1 ring, the MSE with AT loss model reported statistically significant better results than using only CE or MSE ($p < 0.0042$), which is consistent with its behavior in the 3 CMM ($p < 0.0416$). No statistically significant differences in performance were observed at the full volume level (two-tails test, $p > 0.0730$).

\begin{table*}[t]  
\centering
\caption{Volume-wise mean $\pm$ standard deviation Dice values in the test set for each photoreceptor segmentation model and the different areas.}
\resizebox{\textwidth}{!}{
\begin{tabular}{C{4.5cm} | C{1.9cm} | C{1.9cm} | C{1.9cm} | C{1.9cm}}
\hline
\textbf{Method} & \textbf{CSF} & \textbf{3 CMM} & \textbf{3-1 ring} & \textbf{Full volume} \\
\hline
\hline
CE loss &	0.622 $\pm$~0.271	 &	0.691 $\pm$~0.242	 &	0.708 $\pm$~0.242 & 0.820 $\pm$~0.118 \\
\hline
CE + AT loss (CE,~$\omega=8$,~$\lambda_1=1$,~$\lambda_2=8$) & 	\textbf{0.656 $\pm$~0.256}		& \textbf{0.718 $\pm$~0.218}	&	\textbf{0.732 $\pm$~0.218}& \textbf{0.828 $\pm$~0.100} \\
\hline
\hline
MSE loss	& 0.560 $\pm$~0.303 &		0.707 $\pm$~0.223	 &	0.727 $\pm$~0.223 & \textbf{0.835 $\pm$~0.096}\\
\hline
MSE + AT loss (MSE,~$\omega=32$,~$\lambda_1=\lambda_2=1$)	& \textbf{0.708 $\pm$~0.254}	&	\textbf{0.749 $\pm$~0.215}	&	\textbf{0.760 $\pm$~0.213} & 0.821 $\pm$~0.102\\
\hline
\end{tabular}}
\label{tab:table-dice} 
\end{table*}

\begin{figure}[t]
    \centering
    \includegraphics[width=0.98\textwidth]{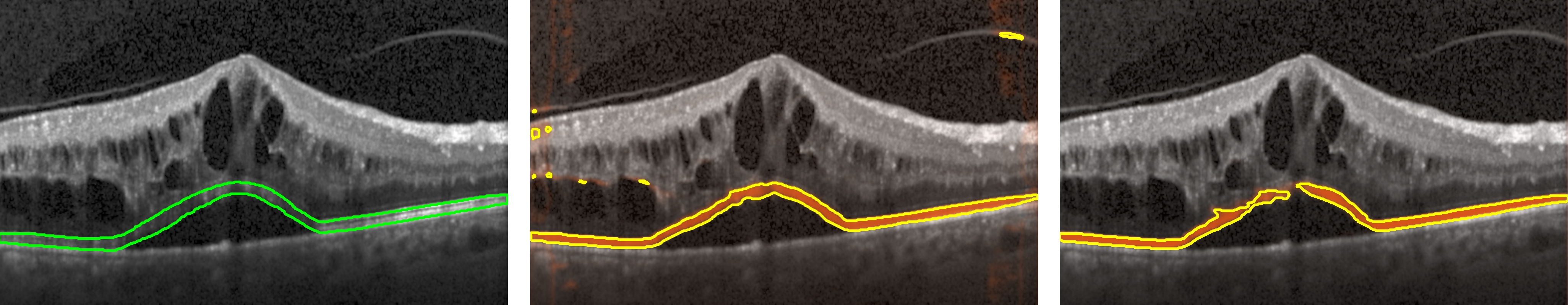}
    \caption{Qualitative effect of the loss selection in the pixel score values. From left to right: manual annotation (green), score map (orange) and binary segmentation (yellow) obtained with MSE + AT loss (MSE, $\omega=32, \lambda_1=\lambda_2=1$) and CE + AT loss (CE, $\omega=8$, $\lambda_1=1$, $\lambda_2=8$).}
    \label{fig:score-maps}
\end{figure}

We qualitatively analyzed the segmentation and score maps  using the CE and MSE combined with the AT loss. Fig.~\ref{fig:score-maps} depicts segmentation results in a central B-scan from the test set, with score maps represented as heatmaps. Using MSE produces noisy scores within the lateral areas of the B-scans, and therefore spurious elements in the segmentation. CE, on the contrary, results in smoother score maps, although with few false negatives in the vicinity of subretinal fluid. This behavior is linked to the one observed in Table~\ref{tab:table-dice}, where the MSE + AT loss model reported higher Dice in the central area than using CE, and smaller values in the full volume. The model trained with only MSE performs poorly in the CSF, the 3 CMM and the 3-1 ring, which indicate that it struggles to deal with pathologies. Similarly, the high performance at a volume level indicates that it can better characterize normal appearances. When using MSE + AT loss, a significant reduction in the amount of false negatives occurs at the central areas. However, as mentioned before, the score maps are noisy at the borders of the B-scans, which causes a drop in the full volume Dice. The model trained with CE + AT loss is less accurate at the center than the one trained with MSE + AT loss, but it still outperforms the baseline approaches. Moreover, at a volume basis the CE + AT loss remains competitive with respect to the one trained only with CE loss.

\begin{figure}[t]
    \centering
    \includegraphics[width=0.98\textwidth]{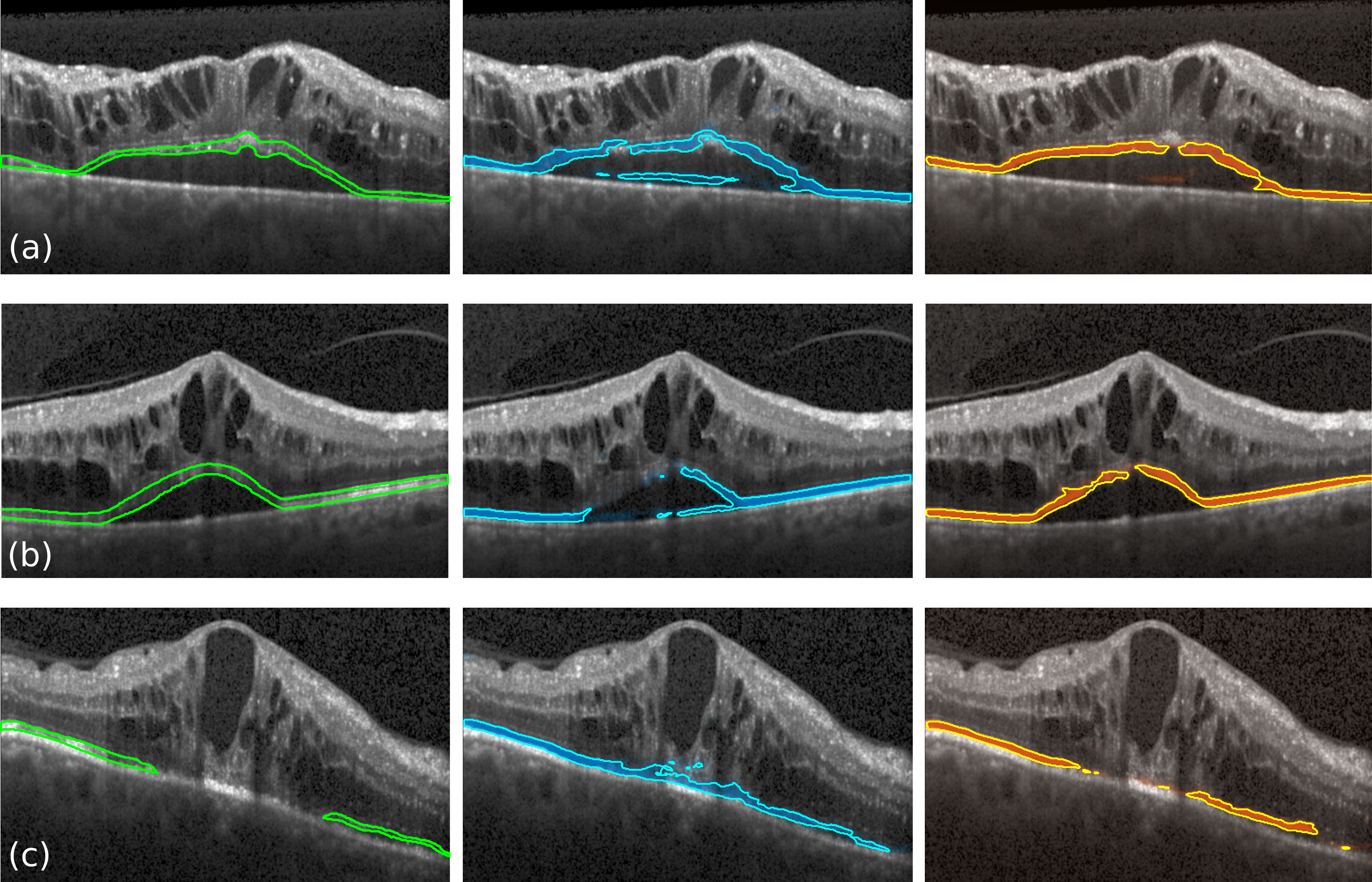}
    \caption{Qualitative results in central B-scans from the test set. From left to right: manual annotations (green), results with only CE loss (blue) and results with CE + AT loss (CE, $\omega=8$, $\lambda_1=1$, $\lambda_2=8$).}
    \label{fig:qualitative-results}
\end{figure}

Finally, Fig.~\ref{fig:qualitative-results} presents qualitative results in exemplary central B-scans from our test set obtained both by the models trained with CE only and  with CE + AT loss. Our approach produced more anatomically plausible segmentations than the standard CE loss in pathological areas with subretinal fluid (Fig.~\ref{fig:qualitative-results} (a) and (b)) or large disruptions (Fig.~\ref{fig:qualitative-results} (c)).

\section{Conclusions}
\label{sec:conclusions}

In this paper we proposed to use the framework of augmented target loss functions for photoreceptor layer segmentation in pathological OCT scans. We define an amplified-target loss incorporating a transformation that weights the central area of the input B-scans to further penalize errors committed in this region. We experimentally observed that this straightforward approach allows to significantly improve performance within the central millimeters of fovea-centered OCT scans, without affecting the overall performance in the entire volume. These results indicate that the proposed AT loss function acts as a form of regularization, better characterizing photoreceptors appearance within highly pathological regions. We are currently exploring new alternatives to identify the regions to weight and to learn their corresponding weights. Further experiments are also performed to evaluate our approach in the context of other OCT based applications such as fluid segmentation and using OCT scans from other vendors.

\noindent\\
\textbf{Acknowledgements}\\
This work is funded by WWTF AugUniWien/FA7464A0249 (MedUniWien); VRG12-009 (UniWien). We thank NVIDIA Corporation for donating a GPU.

\bibliography{paper13.bib}
\bibliographystyle{splncs.bst}

\end{document}